# Predicting Lake Erie Wave Heights using XGBoost

Haoguo Hu[1] and Philip Chu[2]
1. University of Michigan, Ann Arbor  2. NOAA/GLERL

## Abstract

Dangerous large wave put the coastal communities and vessels operating under threats and wave predictions are strongly needed for early warnings. While numerical wave models, such as WAVEWATCH III (WW3, developed at NOAA/NCEP), are useful to provide spatially continuous information to supplement *in situ* observations, however, they often require intensive computational costs. An attractive alternative is machine-learning method, which can potentially provide comparable performance of numerical wave models but only requires a small fraction of computational costs. In this study, we applied and tested a novel machine learning method based on XGBoost for predicting waves in Lake Erie in 2016-2017. To compare the results from the machine learning method, we also conducted numerical wave simulations using WW3 for the same years. XGBoost is an optimized distributed gradient boosting model, it provides a parallel tree boosting that solves data science problems and has been used in a variety of fields. In this study, buoy data from 1994 to 2017 were processed for model training and testing. We trained the model with data from 1994-2015, then used the trained model to predict 2016 and 2017 wave features. The mean absolute error of wave height is about 0.11-0.18 m and the maximum error is 1.14-1.95 m, depending on location and year. For comparison, an unstructured WW3 model was implemented in Lake Erie for simulating wind generated waves. The WW3 results were compared with buoy data from National Data Buoy Center (NDBC) in Lake Erie, the mean absolute error of wave height is about 0.12-0.48 m and the maximum error is about 1.03-2.93 m. The results show that WW3 underestimates wave height spikes during strong wind events and The XGBoost improves prediction on wave height spikes. The XGBoost runs much faster than WW3. For a model year run on a supercomputer, WW3 needs 12 hours with 60 CPUs while XGBoost needs only 10 minutes with 1 CPU. In summary, the XGBoost provided comparable performance for our simulations in Lake Erie wave height and the computational time required was about 0.02 % of the numerical simulations.

## 1. Introduction

Wave conditions are important for ships and coastal regions. Numerous papers and books on wave studies have been published. In the modern computer era, numerical wave models have been developed and implemented in ocean and coastal regions. For Lake Erie, first or second-generation wave model used simple parameterization to account for the nonlinear wave interactions (Schwab et al. 1984). The third-generation wave models that consider three-and-four-wave interactions (SWAN, 2006) have been applied in Lake Erie (Moeini and Etemad-Shahidi 2007). An unstructured coupled FVCOM with SWAN (Qi et al. 2009) was applied in Lake Erie to investigate wave climatology and inter-basin wave interactions (Niu and Xia 2016). WaveWatch III (WW3DG, 2016) is a third generation wave model developed at NOAA/NCEP, which solves the random phase spectral action density balance equation for wavenumber-direction spectra. It was developed for a Great Lakes wave forecasting system (Alves et al. 2014).

In the past15 years, due to Convolutional Neural Network and Recurrent Neural Network and other deep learning algorithms breakthrough, machine/deep learning has been widely used in computer vision, natural language processing, and other scientific study fields. Oceanographers are eager to apply these algorithms for wave studies. Peres et al (2015) used wind (u, v) as input, observed wave heights as output to train a single hidden-layer feed-forward neural network and use the trained model to extend an observed time series of significant wave heights (to repair missing data, Figure 1a). James et al. (2018) used three wave-characteristics: height, period, and direction plus 12x2 wind speeds (u, v), and 357x2 ocean currents (u, v), total 741 input variables, wave height from SWAN model as output to train a neural model as a surrogate model to estimate ocean-wave conditions (Figure 1b).

We used buoy data to train a new deep learning model, XGboost (Chen and Guestrin 2016), then used the trained model to predict wave conditions. XGBoost is an optimized distributed gradient boosting library designed to be highly efficient, flexible, and portable. It implements machine-learning algorithms under the Boosting trees. XGBoost provides a parallel tree boosting that solves many data science problems in a fast and accurate way (Figure 1c).

We also ran a case as a comparison of the full-spectral third-generation wind-wave model WW3 for wave study. The WW3 model is described in the user manual and system documentation (WW3DG, 2016). Model results were validated using buoy data from the NDBC. In this study, an unstructured WW3 was applied in Lake Erie to investigate wave conditions at stations 45005 and 45142.

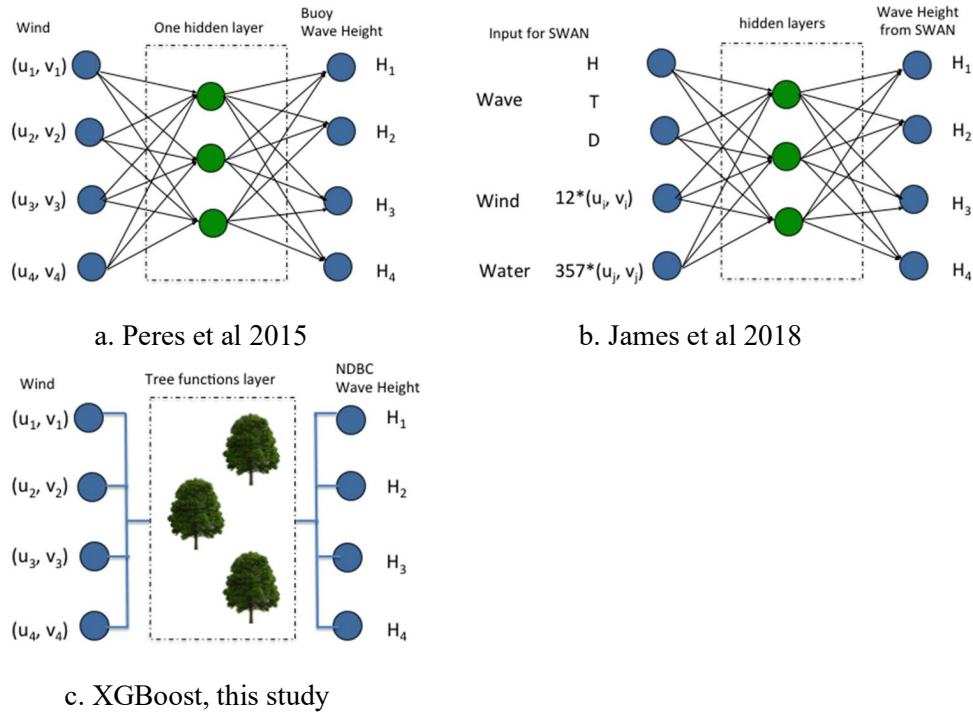

a. Peres et al 2015

b. James et al 2018

c. XGBoost, this study

**Figure 1.** Scheme of machine learning models mentioned in the introduction

## 2. XGBoost Model
### 2.1 Introduction of XGBoost

XGBoost was developed and described in detail by Chen and Guestrin (2016). Here, we give a brief description of the model. XGBoost is mostly used for supervised learning problems, where one can use the training data $x_i$, to predict a target variable $\hat{y}_i$. XGBoost implements machine learning algorithms under the Gradient Boosting framework, it uses the same tree ensemble model as decision tree and random forest models. The tree ensemble model consists of a set of classification and regression trees. Tree ensemble is also the model used in random forests. Basically, random forests and boosted trees are the same models; the only difference is how one trains them.

Mathematically, Tree model can be written in the form
$\hat{y}_i = \sum_{k=1}^{K} f_k(x_i), f_k \in F$ Where K is the number of trees, f(x) is a tree function in the function space F, and F is space of functions containing all regression trees.

The objective/loss function with parameters $\theta$ to be optimized is given by
$L(\theta) = \sum_i l(y_i, \hat{y}_i) + \sum_k \Omega(f_k)$
Where $l$ is a differentiable convex loss function that measures the difference between the prediction $\hat{y}_i$ and the target $y_i$. $\Omega(f)$ is the regularization term, which penalizes the complexity of the model.

Let $\hat{y}_i^{t-1}$ and $\hat{y}_i^t$ be the predictions at time (t-1) and t iteration, respectively, if $\hat{y}_i^t = \hat{y}_i^{t-1} + f_t(x_i)$, then the loss function at t-th iteration will be:
$L^t = \sum_i l(y_i, \hat{y}_i^{t-1} + f_t(x_i)) + \Omega(f_t)$
One has to decide to add what kind of $f_t(x)$ to minimize the loss function L. Recall Taylor expansion $f(x + \Delta x) = f(x) + f'(x)\Delta x + \frac{1}{2} f''(x)\Delta x^2$ then

$$l(y_i, \hat{y}_i^{t-1} + f_t(x_i)) = l(y_i, \hat{y}_i^{t-1}) + \frac{\partial l(y_i, \hat{y}_i^{t-1})}{\partial \hat{y}_i^{t-1}} f_t(x_i) + \frac{1}{2} \frac{\partial^2 l(y_i, \hat{y}_i^{t-1})}{\partial \hat{y}_i^{t-1} \partial \hat{y}_i^{t-1}} f_t^2(x_i)$$

Then we have
$$L^t \cong \sum_i [l(y_i, \hat{y}_i^{t-1}) + g_i f_i(x_i) + \frac{1}{2} h_i f_i^2(x_i)] + \Omega(f_t) + constant$$
where $g_i = \partial_{\hat{y}_i^{t-1}} l(y_i, \hat{y}_i^{t-1})$ and $h_i = \partial_{\hat{y}_i^{t-1}}^2 l(y_i, \hat{y}_i^{t-1})$.

Define complexity $\Omega(f) = \gamma T + 0.5\lambda \sum_{j=1}^{T} w_j^2$ where T is the number of leaves in the tree. Then with constant terms removed, the simplified objective (loss function) at step t is:

$$L^t \cong \sum_i [g_i f_i(x_i) + \frac{1}{2} h_i f_i^2(x_i)] + \gamma T + 0.5\lambda \sum_{j=1}^{T} w_j^2$$

Define the instance set in leaf j as $I_j = \{i | q(x_i) = j\}$, $f_t(x_i) = w_q(x_i)$ then

$$L^t \cong \sum_i \left[\left(\sum_{i \in I_j} g_i\right) w_j + \frac{1}{2}\left(\sum_{i \in I_j} h_i + \lambda\right) w_j^2\right] + \gamma T$$

Define $G_j = \sum_{i \in I_j} g_i$ and $H_j = \sum_{i \in I_j} h_i$ then

$$L^t \cong \sum_i [G_j w_j + \frac{1}{2}(H_j + \lambda) w_j^2] + \gamma T$$

And then $\quad argmin_x(L) = -\frac{1}{2} \sum_{j=1}^{T} \frac{G_j^2}{H_j + \lambda} + \gamma T \quad$ when $w^* = -\frac{G_j^2}{H_j + \lambda}$

In summary, the boosted tree algorithm:

1. Adds a new tree in each iteration
2. At the beginning of each iteration, calculates: $g_i = \partial_{\hat{y}_i^{t-1}} l(y_i, \hat{y}_i^{t-1})$ and $h_i = \partial^2_{\hat{y}_i^{t-1}} l(y_i, \hat{y}_i^{t-1})$
3. Use the statistics to greedily grow a tree $f_t(x)$ $\quad L = -\frac{1}{2} \sum_{j=1}^{T} \frac{G_j^2}{H_j + \lambda} + \gamma T$
4. Add $f(x)$ to the model $\quad \hat{y}_i^t = \hat{y}_i^{t-1} + \varepsilon f_t(x_i)$ where $\varepsilon$ is called shrinkage

## 2.2 Configurations and runs for XGBoost

The model was trained with 1994-2015 hourly wind and wave height data of Station-45005 and Station-45142 from the National Data Buoy Center. The data from the stations were used to train the model separately. The wind speeds and directions were used as input $X_i$, and wave heights were used as model output $y_i$. We used 60% of the data set for training and 40% of the data set for validating. Other parameters are: objective = reg:linear, eval_metric = mae, max-depth = 5. The model ran on a CPU with about 10 minutes and it converged at around 210 epochs.

The trained model was used to predict wind-wave conditions for 2016-2017. For the model trained by Station-45005 data, it had mean absolute errors of 0.112 m and 0.113 m; maximum absolute error of 1.146 m and 1.142 m for 2016 and 2017, respectively. For the model trained by Station-45142 data, it had mean absolute errors of 0.155 m and 0.175 m; the maximum absolute error of 1.737 m and 1.946 m for 2016 and 2017, respectively. The model predicted results were compared to NDBC buoy data in **Figure** 2 and **Figure 5**. The model produced noticeably better results for strong wind events (spikes on the figures) when compared to WW3.

What about if the model was trained by Station-45005 data and use to predict Station-45142 wave conditions, and vice versa? We used the trained model to make predictions for 2017. The results are shown in **Figure 3**, **Figure 4**, and **Figure 5**. It turned out that the errors are 30-40% larger than the previous one.

We also made a test to train the model with 6 years of data (2010-2015); the trained model was used to predict wind-wave conditions for 2016-2017. For Buoy station 45005, the mean absolute error was 0.117 m and 0.121 m; the maximum absolute error was 1.093 m and 1.200 m for 2016 and 2017, respectively. For Buoy station 45142, the mean absolute error was 0.180 m and 0.203 m; the maximum absolute error was 2.217 m and 1.901 m for 2016 and 2017, respectively. The comparisons of the model predictions to NDBC buoy data are shown in **Figure 5**.

In summary, models trained with more data tend to perform better. Model results are sensitive to locations, so it is better to train individual models according to locations for better accuracy.

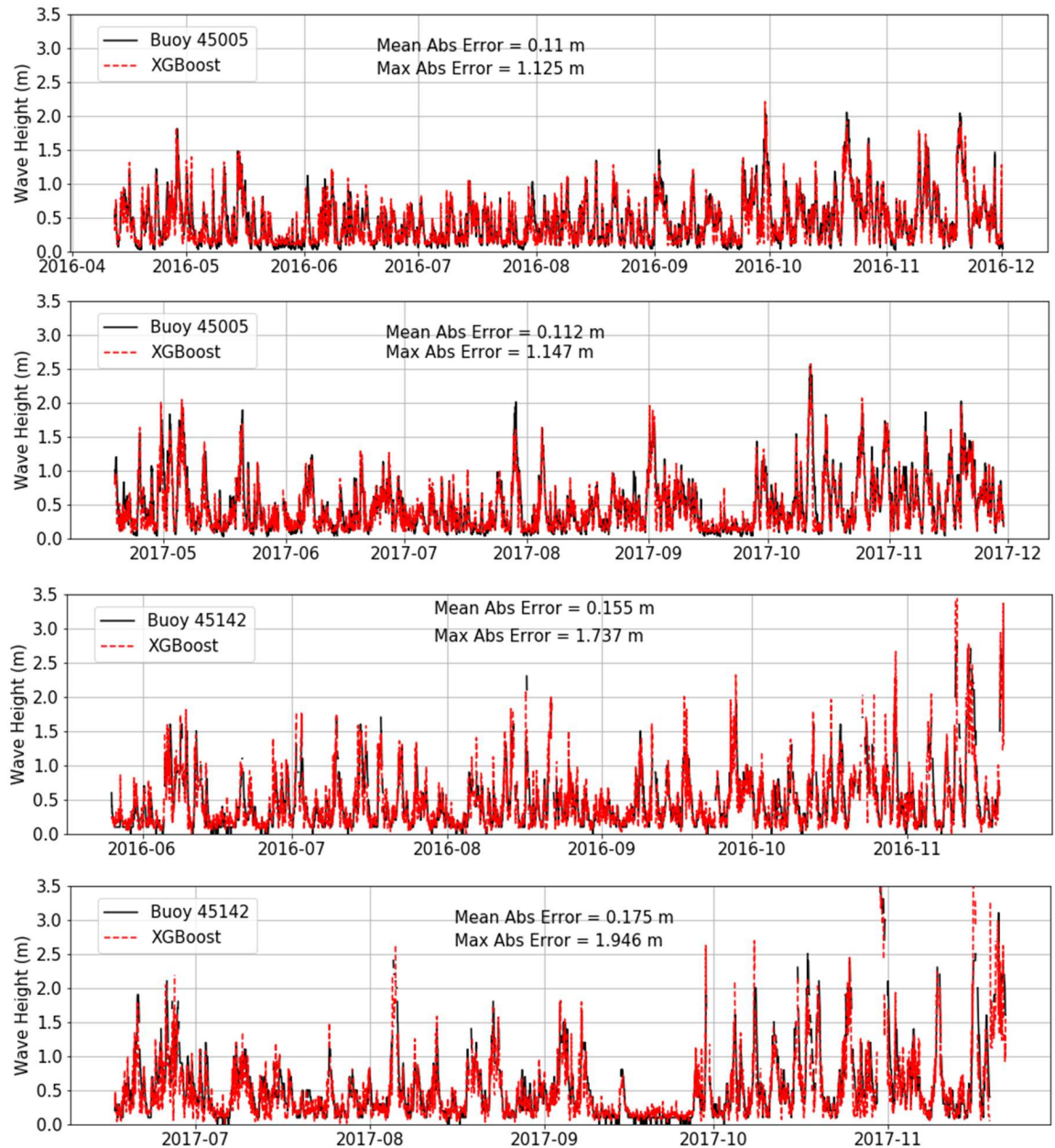

**Figure 2.** Model was trained with 1994-2015 buoy data, and then the trained model was used to predict 2016-2017 wave heights.

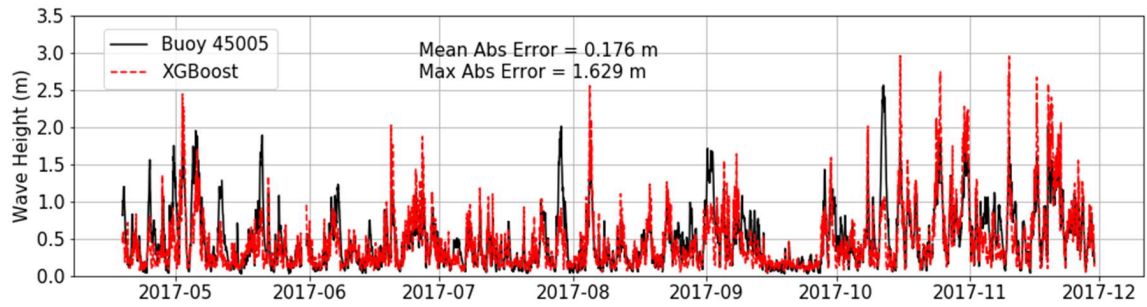

**Figure 3.** Model was trained with Buoy 45142 data (22 years), and then the trained model was used to predict wave height (2017) for station 45005

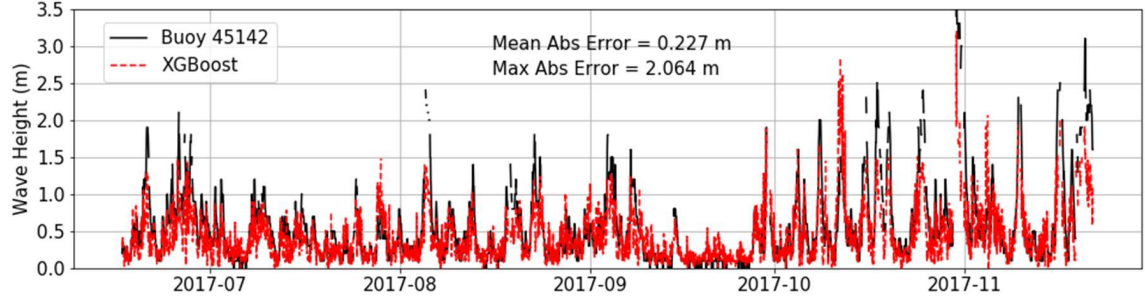

**Figure 4.** Model was trained with Buoy 45005 data (22 years), and then the trained model was used to predict wave height (2017) for station 45142

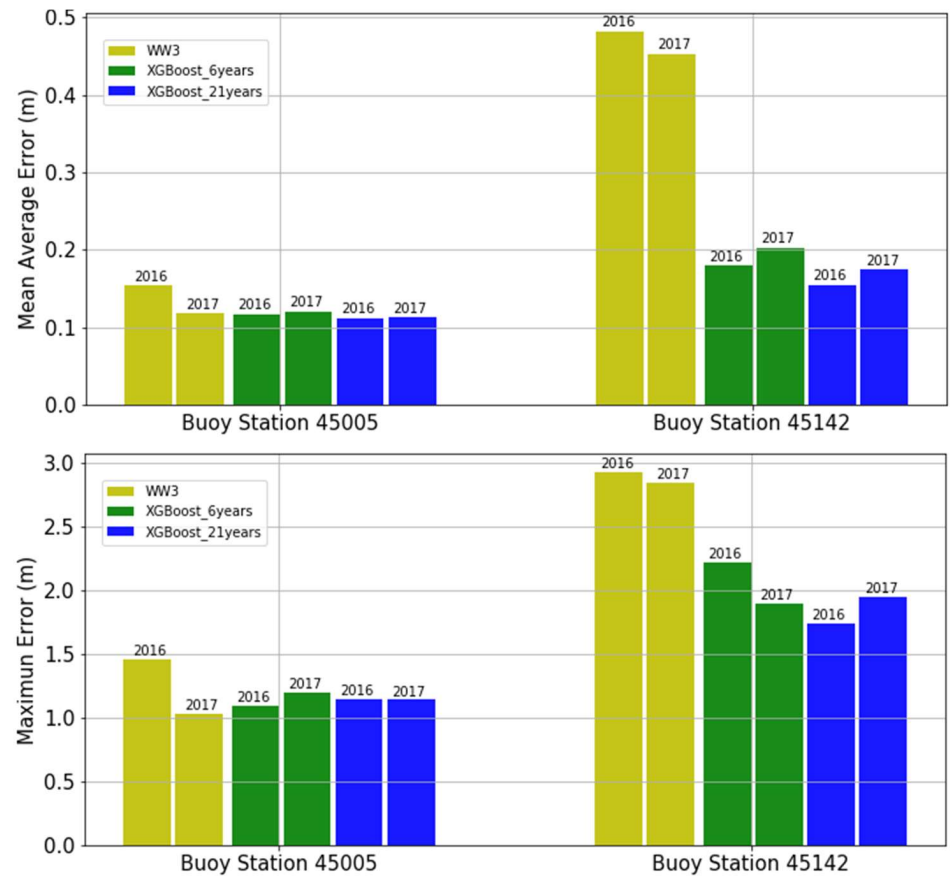

**Figure 5.** Performances of WW3 and XGBoost. Mean average error is shown on top panel, Maximum error is shown on bottom panel.

## 3. Configurations and runs for WaveWatch III

The grid generation module of the Surface-Water-Modeling System software was used to generate the unstructured model grid. The grid size distribution is configured as dependent on the bathymetry (NOAA NGDC, 3 arc-second). The model bathymetry was obtained by interpolating the bathymetry onto each unstructured grid node. The model grid is shown in **Figure** 6, high-resolution NOAA coastline data was applied to delineate the land boundary. The model grid in the horizontal is composed of 11,509 triangular elements and 6,106 nodes. The resolution varies from approximately 100 m near the shore to 2.5 km offshore. The model has distribution referenced to the Great Lakes low water datum of 173.5 m.

Surface wind observations were interpolated to create hourly gridded surface meteorological analyses of wind. The observations are from both land-based stations such as ASOS and AWOS stations, coastal stations including NWS/NDBC C-MAN stations, ECCC automated stations on lighthouses and offshore platforms, NOS/CO-OPS NWLON meteorological stations, and Other Marine Reports.

The model simulation starts at 12:00 GMT on 01 Jan 2016. Model results are output on a 1-hour time step at the same time step of buoy data. The model runs around 12 hours on 60 CPUs for a model year. The switches used in this model are "F90 NC4 NOGRB NOPA LRB4 DIST MPI PR3 UQ FLX0 LN1 ST4 STAB0 NL3 BT1 DB1 MLIM TR0 BS0 IC0 REF0 XX0 WNT1 WNX1 CRT0 CRX0 IS0 O0 O1 O2 O4 O5 O6 O7 O14 O15"

The model was implemented to predict wind-wave conditions for 2016-2017. For Station 45005, it had mean absolute errors of 0.15 m and 0.12 m, and maximum absolute errors of 1.46 m and 1.03 m for 2016 and 2017, respectively. For Station 45142, it had mean absolute errors of 0.48 m and 0.45 m; the maximum absolute error of 2.93 m and 2.84 m for 2016 and 2017, respectively. The model predicted results were compared to NDBC buoy data in **Figure** 7 and Table 1. The model tends to underestimate wave height peaks during strong wind events (spikes on the figures).

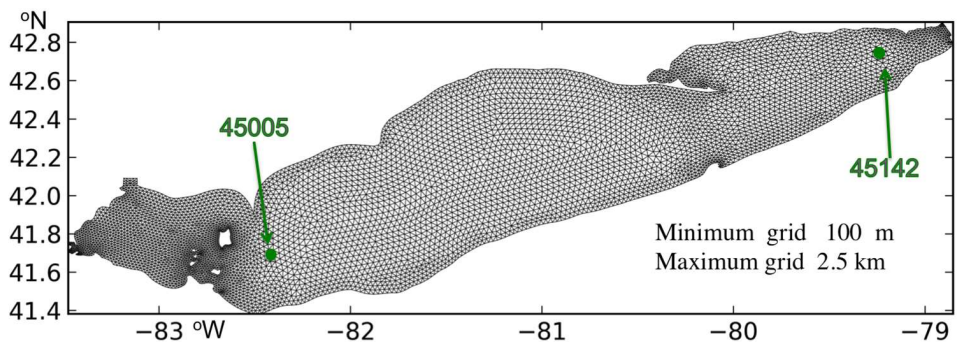

**Figure 6.** Unstructured grid of the Lake Erie used for WW3. The buoy stations of 45005 and 45142 are also labeled on the figure.

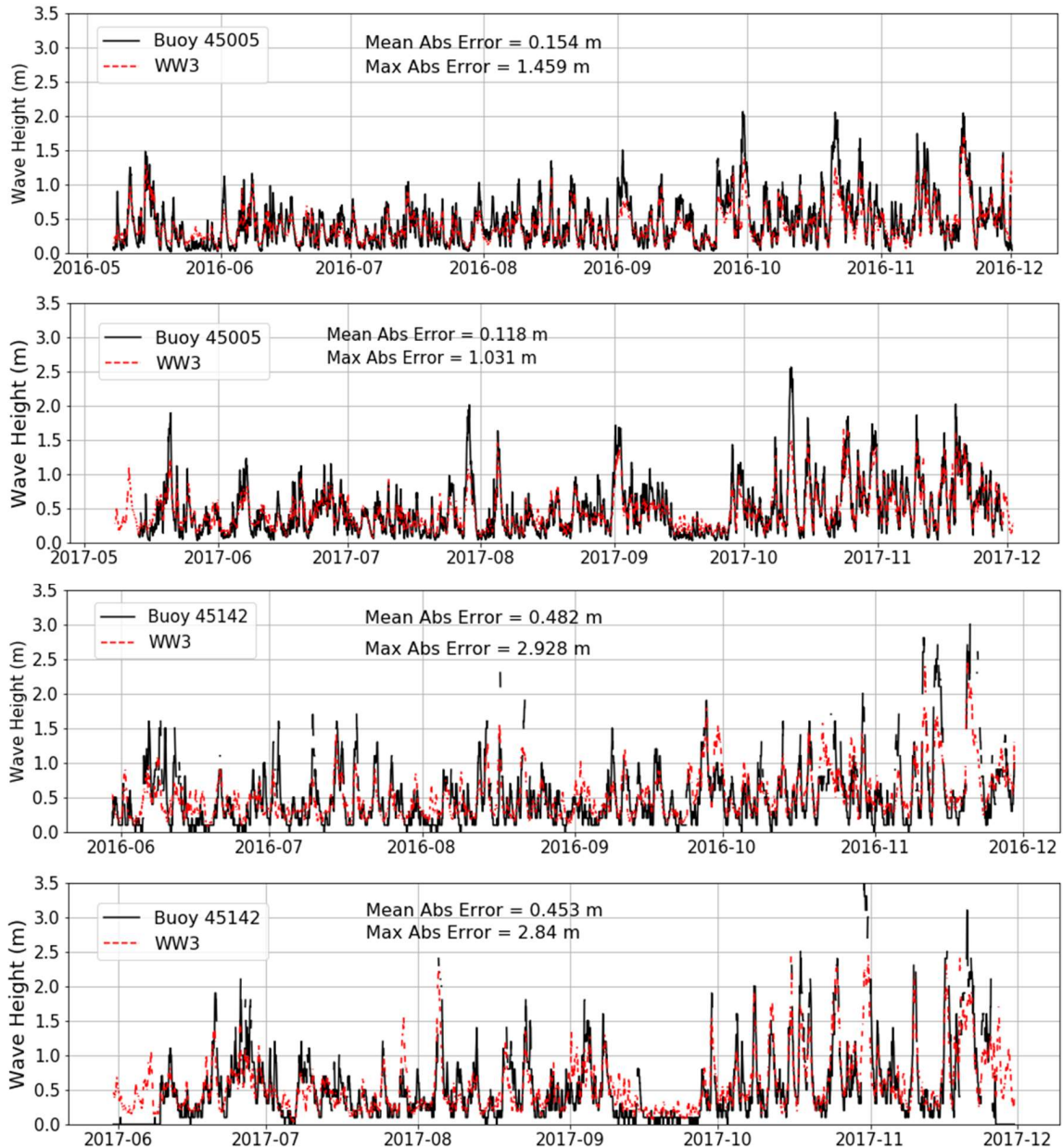

**Figure 7.** WW3 predicted 2016-2017 wave heights for stations 45005 and 45142.

## 4. Conclusions

A novel machine learning method XGBoost is applied to wind-wave prediction in Lake Erie. Buoy data from 1994 to 2017 were processed for model training and testing. We used 1994-2015 data for training the model, and used the trained model to predict 2016 and 2017 wave features. The mean absolute error of wave height was 0.11-0.18 m with a maximum error of 1.14-1.95 m, depending on location and year. Results show that a model trained with more data tends to have better performance. XGBoost is sensitive to locations, so it is better to train individual model according to locations to gain better accuracy. As a comparison, an unstructured WW3 model was also implemented in the Lake Erie for simulating wind generated waves. The WW3 results were compared with buoy data in Lake Erie. The mean absolute error of wave height is 0.12-0.48 m and the maximum error is 1.03-2.93 m, depending on location and year. WW3 basically underestimates wave height spikes during strong wind events and XGBoost improves prediction on those events. XGBoost provided better performance for our simulations in Lake Erie and the computational time required was only about 0.02 % of the numerical simulations. It should be mentioned that XGBoost outputs only wave height in this study; whereas WW3 is capable of outputting wave height, period, directions, and wave spectrums.